\documentclass[10pt,aps,prl,amsfonts,amssymb,floatfix,superscriptaddress,notitlepage,twocolumn,footinbib]{revtex4-1}
\usepackage{mathrsfs} 
\usepackage{graphicx,color} \usepackage{bbm}
\usepackage{bm}
\usepackage[normalem]{ulem}
\usepackage{physics}
\usepackage{afterpage}

\def\to{\rightarrow}  

\newcommand{\blue}[1]{\textcolor{black}{#1}}
\newcommand{\red}[1]{\textcolor{black}{#1}}
\newcommand{\eq}[1]{Eq.~(\ref{#1})}	
\newcommand{\figcaption}[1]{\def\@captype{figure}\caption{#1}}
\newcommand{\tblcaption}[1]{\def\@captype{table}\caption{#1}}
\makeatother

\newcounter{num}

\newcommand{\ctext}[1]{\raise0.1ex\hbox{\scriptsize \textcircled{\tiny {#1}}}}
\newcommand{\cctext}[1]{\raise0.1ex\hbox{\textcircled{\scriptsize {#1}}}}

\newcommand{\beq}{\begin{equation}}
\newcommand{\eeq}{\end{equation}}

\begin{document}

\title{Spin-orbital glass transition in a model of a frustrated pyrochlore magnet without quenched disorder}

\author{Kota Mitsumoto}
\affiliation{Graduate School of Science, Osaka University, Toyonaka, Osaka 560-0043, Japan}

\author{Chisa Hotta}
\affiliation{Department of Basic Science, University of Tokyo, Tokyo 153-8902, Japan}

\author{Hajime Yoshino}
\affiliation{Cybermedia Center, Osaka University, Toyonaka, Osaka 560-0043, Japan}
\affiliation{Graduate School of Science, Osaka University, Toyonaka, Osaka 560-0043, Japan}

\begin{abstract}
We show theoretically that spin and orbital degrees of freedom 
in the pyrochlore oxide Y$_2$Mo$_2$O$_7$, which is free of quenched disorder, 
can exhibit a simultaneous glass transition, working as dynamical
disorder for each other. 
The interplay of spins and orbitals is mediated by the Jahn-Teller lattice distortion that selects the
choice of orbitals, which then generates variant spin exchange interactions ranging from ferromagnetic to antiferromagnetic ones. 
Our Monte Carlo simulations detect the power-law divergence of the relaxation times 
and the negative divergence of both the magnetic and dielectric nonlinear susceptibilities, 
resolving the long-standing puzzle on the origin of the disorder-free spin glass. 
\end{abstract}

\maketitle
\red{Glass formation is a generic phenomenon that emerges in strongly frustrated systems \cite{tarjus2005frustration},
ranging from super-cooled molecular liquids and densely packed soft-matters
 \cite{angell2000relaxation,sciortino2005glassy,cavagna2009supercooled,berthier2011theoretical}
to geometrically frustrated spin systems on
kagome  \cite{martinez1992magnetic,ladieu2004relative,hamp2018supercooling,cugliandolo2019mean,upadhyay2017magnetic} and 
pyrhoclore lattices  \cite{alba1982very,greedan1986spin,reimers1991short,gingras1996nonlinear,gingras1997static,gardner1999glassy,zhou2008unconventional,taniguchi2009spin,ladieu2004relative,thygesen2017orbital,booth2000local}.
Frustration suppresses ordinary long-range orders and stabilizes
liquid/paramagnetic states down to unusually low temperatures where
the effect of many-body interactions become dominant.
In sharp contrast to normal liquids at higher temperatures,
molecules in super-cooled liquids can flow only through collective motions
because of the strong inter-molecular interactions.}
In classical antiferromagnets on triangular, kagom\'{e} and pyrochlore lattices, 
the frustration among interactions caused by the lattice geometry precludes
the possibility of all bonds to be simultaneously satisfied in energy \cite{diep2013frustrated}.
\red{An outstanding open question there is whether truly thermodynamic glass transitions, for example the putative Kauzmann transition of super-cooled liquids \cite{Ka48}, are possible.
At present such ideal glass transitions is attained only theoretically in unphysical limits of infinitely large dimensions for some super-cooled liquids  \cite{parisi2010mean,kurchan2012exact,charbonneau2014fractal}
or spin systems with infinitely large number of spin components  \cite{yoshino2018disorder}.}

\red{In  realistic three-dimension, a true thermodynamic glass transition is established only in lattice systems with quenched disorder added on top of the frustration  \cite{mydosh2014spin,binder1986spin,kawamura2015spin}. Theoretically, this spin glass(SG) is found  in the Edwards-Anderson model  \cite{edwards1975theory,mezard1987spin} based on both Ising  \cite{ogielski1985dynamics,bhatt1988numerical,kawashima1996phase} and Heisenberg spins  \cite{hukushima2000chiral,lee2007large,viet2009numerical,ogawa2019monte}. }

\red{The next fundamental question is whether the SG transition can be sustained even when such quenched disorder is switched off. 
It is known experimentally that some of the disorder-free pyrochlore magnets exhibit a sharp SG transition \cite{alba1982very,greedan1986spin,reimers1991short,gingras1996nonlinear,gingras1997static,gardner1999glassy,zhou2008unconventional,taniguchi2009spin,ladieu2004relative},} which is indistinguishable from the conventional one with quenched disorder  \cite{mydosh2014spin,binder1986spin,kawamura2015spin}. \red{The divergence of nonlinear susceptibility  \cite{gingras1996nonlinear} establishes this phase as SG. In contrast,, any of the theories available at present does not allow for the 3D SG without quenched disorder. }

This Letter aims to clarify the origin of the SG transition in the disorder-free pyrochlore magnets 
by constructing a realistic model consisting of \red{two key degrees of freedom of the system, i.~e.
spins and orbitals which are coupled through a 
Jahn-Teller (JT) lattice distortion which modifies the super-exchange interactions between the spins dynamically.
The two degrees of freedom, if decoupled, live in highly degenerate energy landscapes of their own, due to the strong geometrical frustration,
so that they remain in their liquid/paramagnetic states at all temperatures.
Through the coupling they correlate producing
dynamical disorder for each other 
and create an emergent lugged free-energy landscape.
Our extensive  MC simulation uncovers an unprecedented  thermodynamic glass transition where the two degrees of freedom simultaneously freeze cooperatively.}
\begin{figure}[t]
\centering
\includegraphics[width=80mm]{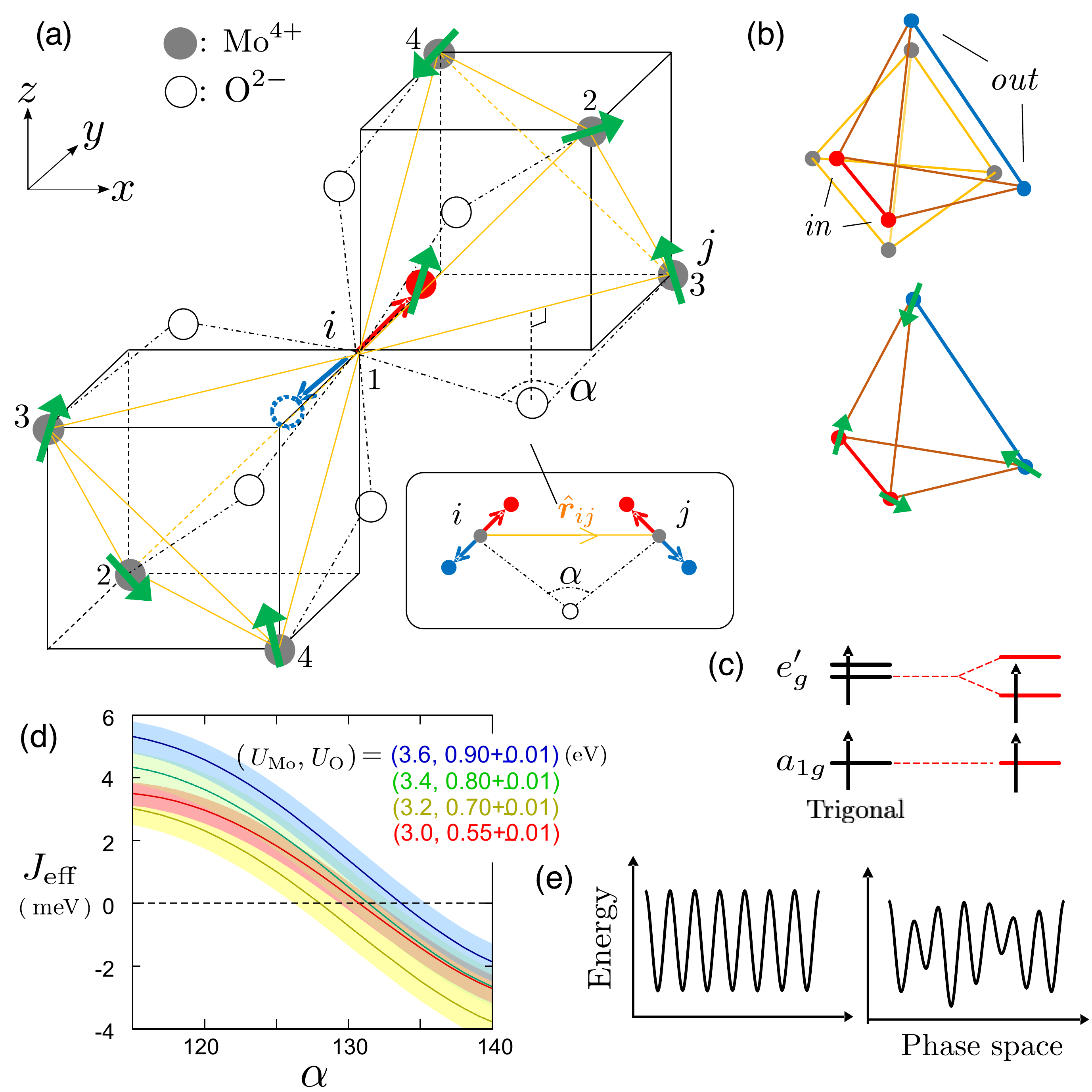}
\caption{
  (a): O$^{2-}$ ions and magnetic Mo$^{4+}$ ions around the $i$ site, where the numbers 1-4 are the sublattice indices of Mo$^{4+}$ ions.
  Red and blue dashed circles represent the positions of the Jahn-Teller distorted $i$ ion.
  Spins on a pair of Mo$^{4+}$ ions $(i,j)$ interact through the O$^{2-}$ ion as shown in the inset panel, where alpha is the Mo-O-Mo angle and $\hat{r}_{ij}$ is the unit vector in the $i \rightarrow j$ direction. 
  (b): Ice-type displacements of the Mo tetrahedron. The different color bonds represent different exchange interactions.
  \red{
(c): Lifting of the orbital degeneracy of $e_g'$ by the JT effect.
  (d): $\alpha$-dependence of the effective exchange interaction for typical on-site Coulomb interactions on Mo and O ions
  $U_{\rm MO}$  and $U_{\rm O}$ (see SI). The shaded regions are the variance of
      $J_{\rm eff}$ when $U_{\rm O}$ is varied in the range of $\pm 0.01$.}
(e): Schematic pictures of the energy landscape of ice-type displacements (Top) and modified one by
    the coupling between the spin and lattice distortion.
}
\label{Fig1}
\end{figure}

The prototypical materials for our theory are the pyrochlore oxides, $A_2$Mo$_2$O$_7$ 
($A=$ Ho, Y, Dy, Tb), which are insulating and show a SG transition at around 20 K \cite{gaulin1992spin,dunsiger1996muon,gingras1997static,gardner1999glassy,kezsmarki2004charge,hanasaki2007nature}. 
The magnetic ions Mo$^{4+}(4d^{2},S=1)$ sit on the vertices of the corner shared tetrahedra (see Fig. \ref{Fig1}(a)), 
and the interactions between them are antiferromagnetic, 
whose underlying microscopic mechanism is the superexchange mediated by the O$^{2-}$ ions \cite{solovyev2003effects}. 
It has been established that the classical spin model with purely antiferromagneic nearest neighbor interactions 
on the pyrochlore lattice remains nonmagnetic down to zero temperature because of the strong frustration effect \cite{reimers1991mean, moessner1998properties}, not providing any hint of the SG transitions observed in experiments. 

Very recently, some salient features of the local lattice distortions in Y$_2$Mo$_2$O$_7$
 \cite{thygesen2017orbital} were uncovered experimentally. 
The analysis based on the neutron pair-distribution function data 
suggests that the Mo$^{4+}$ ions are locally displaced towards (\textit{in}) 
or away from (\textit{out}) the centers of Mo$_4$ tetrahedra (see Fig. 1(a)), 
explaining well a large variance of the Mo-Mo distances observed in the x-ray absorption study \cite{booth2000local}. 
A natural mechanism of distortion is the JT effect that lowers the electrostatic energy of the Mo ion lifting its orbital degeneracy
\blue{(Fig.\ref{Fig1}(c)) (See SI)}.

The displacements of the four Mo$^{4+}$ ions on each tetrahedron possibly follow the 2-in-2-out rule, 
i. e. the ice-rule (see Fig. 1(a,b)), 
generating huge numbers of ionic configurations over the system equivalent in their JT energies. 
Such distortions thus do not lead to any long-range structural ordering. 

It is empirically known 
\footnote{The evaluation of the direct exchange interactions between Mo spins 
can be done 
based on the Hartree-Fock approach by assuming the orbital ordering, which gives a rough estimate of the sign of the interactions. 
We give a more detailed analysis by considering the dominant superexchange interactions for the insulating Y$_2$Mo$_2$O$_7$.} 
that the exchange interaction may vary significantly and even changes its sign, depending on the degree of the Mo-O-Mo angle $\alpha$
 \cite{solovyev2003effects}. 
Indeed, the angle evaluated from the variance of the displacements of the Mo ions 
distributes in the range $\alpha = 116^\circ$ and $139^\circ$, 
which is large enough to change the sign of the interaction
\blue{(Fig.\ref{Fig1}(d)) (See SI)}.
Then, depending on the configurations of the local lattice distortion in space, 
the magnetic exchange interaction may acquire a random distribution with finite variances. 
This can be thought of as follows; 
If we consider the Jahn-Teller Hamiltonian including only the lattice degrees of freedom, 
the energy landscape should be such that all the ice-rules give the 
energy minima of the same height(see Fig.1(d)) \cite{pauling1960nature, bramwell2001spin}. 
In putting spins on the ice-type displaced vertices, 
its energy landscape is modified to an irregular one with random valley structures, 
since the variance of the spin interactions from ferro to antiferromagnetic ones 
add different energies to each ice configuration. 
We verify the above picture showing numerically that the spin and lattice distortions 
simultaneously undergo a glass transition and freeze into the aperiodic, irregular types of configuration. 
Since the randomly frozen lattice distortions indicate the randomly selected orbital configurations, 
we call it a {\it spin-orbital glass}.

As a natural description of the material faithful to the experimental observations, 
we introduce a disorder-free spin model including not only the classical Heisenberg spins $\bm{S}_i~(i = 1,2,3,...,N)$
for the magnetic moments of the Mo$^{4+}(4d^{2},S=1)$ ions
but also the degree of lattice distortion of the Mo ions $\bm{\sigma}_i$ represented as vectors taking two discrete values, 
$\bm{\sigma}_i = \sigma_i\hat{\bm{e}}_\nu$. 
Here, $\sigma_i=\pm 1$ corresponds to either {\it in} or {\it out} depending on the direction of $\bm e_\nu$ which is the unit vector in the $[111],[1\bar{1}\bar{1}],[\bar{1}1\bar{1}]$ and $[\bar{1}\bar{1}1]$  directions, respectively for the sub-lattices $\nu = 1,2,3,4$ which the $i$-th spin belongs to (See Fig.~1.(a)).  
The Hamiltonian is given by 
\begin{equation}
H = \sum_{<ij>}J_{\bm{\sigma}_i,\bm{\sigma}_j}\bm{S}_i\cdot\bm{S}_j - \epsilon\sum_{<ij>}\bm{\sigma}_i\cdot\bm{\sigma}_j~(\epsilon > 0)
\label{hamiltonian}
\end{equation}
with
\begin{equation}
J_{\bm{\sigma}_i,\bm{\sigma}_j} = J[1+\delta(\bm{\hat{r}}_{ij}\cdot\bm{\sigma}_i + (-\bm{\hat{r}}_{ij})\cdot\bm{\sigma}_j)]~(\delta > 0),
\label{coupling}
\end{equation}
The first term in \eq{hamiltonian} represents the exchange interaction whose coupling constant $J_{\bm{\sigma}_i,\bm{\sigma}_j}$ depends on the angle of Mo-O-Mo bond.
\red{This form is constructed in a way to reproduce the values of $J_{\rm eff}$ derived microscopically from the perturbation process on the Kanamori-type of Hamiltonian \blue{(see SI)}. 
As shown in Fig. \ref{Fig1} (d), for realistic values of the on-site Coulomb interactions on Mo$^{4+}$ and O$^{2-}$ ions  \cite{solovyev2003effects,shinaoka2013spin}, $J_{\rm eff}$ changes its sign within the experimentally observed range of the Mo-O-Mo angle $\alpha = 116^\circ$-$139^\circ$. 
This is modeled by \eq{coupling} as} such that 
$
J_{\bm{\sigma}_i,\bm{\sigma}_j} =
 (1+2\tilde{\delta})J\mbox{ (\textit{in}, \textit{in})}, J \mbox{ (\textit{in}, \textit{out})}, (1-2\tilde{\delta})J \mbox{ (\textit{out}, \textit{out})},
$
where $\tilde{\delta} = \sqrt{6}\delta/3$ (See Fig.~1.(b)).  
The second term of \eq{hamiltonian} represents the elastic energy of the Mo$^{4+}$ displacement. 
The elastic energy is minimized if a Mo$^{4+}$ tetrahedron satisfies the ice-rule. 
More precisely, the elastic energy of each bond takes $\epsilon \bm{\sigma}_i\cdot\bm{\sigma}_j=-\epsilon/3$ 
if both displacements are \textit{out} or \textit{in}, otherwise $+\epsilon /3$. 
\footnote{
This term is the simplified description of the energies of the classical elastic model whose 
lattice sites are connected by springs, for which one can easily find that the two-in two-out configurations 
give the lowest energy, consistent with the experimental findings. 
}
There are three parameters in this system; 
$\tilde{T}=k_{\rm B}T/J$ is the dimensionless temperature, 
$\tilde{\epsilon}=\epsilon/3J$ the ratio of the energy scales between 
the exchange interaction and the elastic energy of the displacement, 
and $\tilde{\delta}=\sqrt{6}\delta/3$ is the amplitude of the displacement 
(hereafter we call them simply as $T,~\epsilon,~\delta$). 
At $\epsilon \to \infty$, the lattice distortion becomes static. 
In the following analysis, we mainly focus on a representative system at $\delta = 1.5, \epsilon = 0.6$.

We consider the periodic systems of cubic geometry consisting of $L^3$ unit cells with totally $N=16L^3$ spins, 
and perform 120 statistically independent runs for the system size $L = 4,5,6,8$, 
evaluating the averages and mean-squared errors of observable. 
To equilibrate the system we made a combined use of the exchange Monte Carlo method \cite{hukushima1996exchange, hukushima1999domain}, conventional and loop \cite{melko2001long} update for lattice variables, Metropolis-Reflection \cite{pawig1998monte} and over-relaxation \cite{PhysRevB.53.2537} update for spin variables. \red{However, to study the relaxational dynamics, we switched to single-spin-flip method after the equilibration. (see SI)}
We took $3\times10^7$ Monte Carlo steps for both equilibration and for taking thermal averages 
$\langle ... \rangle_{\rm eq}$. 

\begin{figure}[t]
\includegraphics[clip,width=85mm]{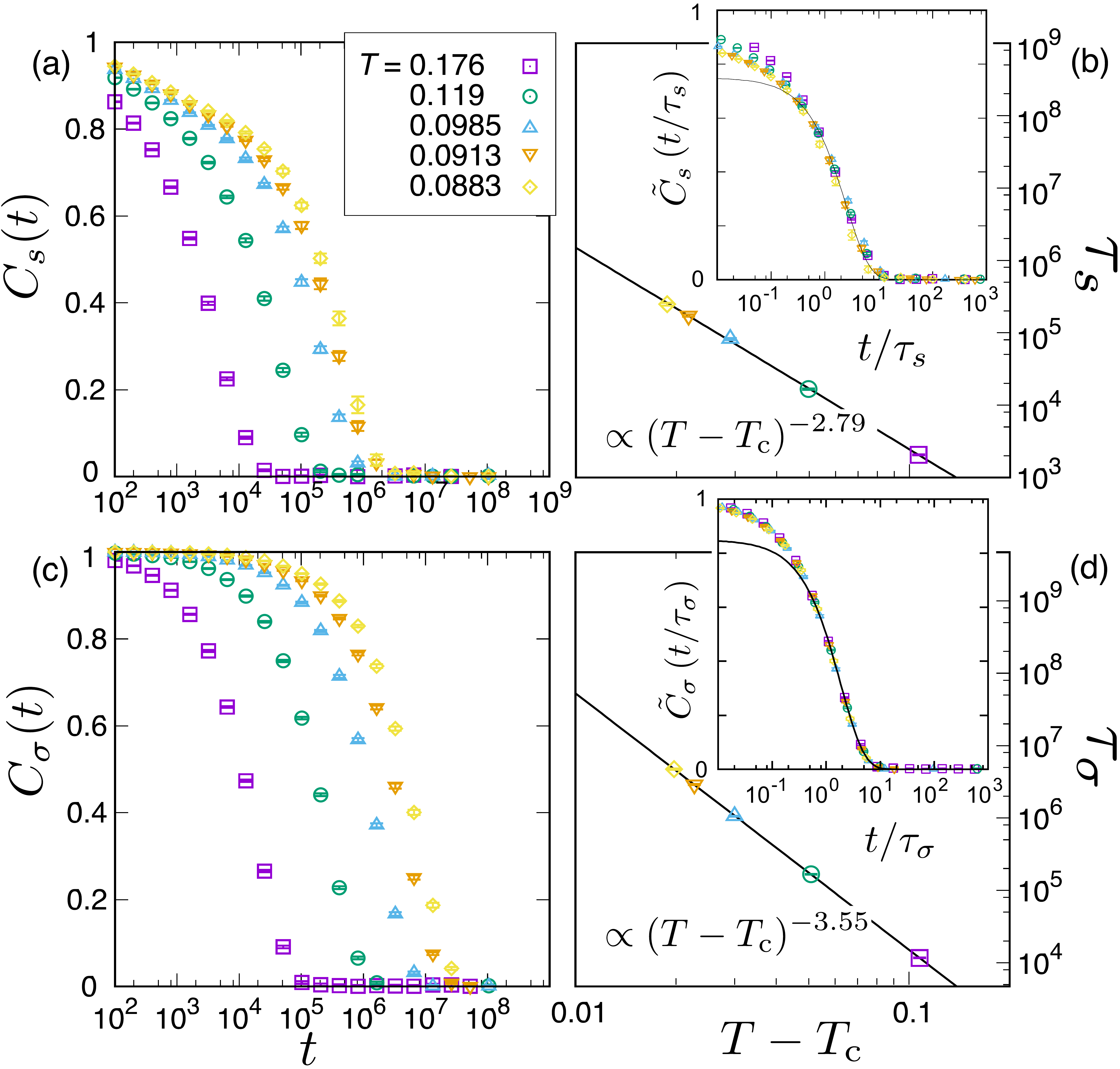}
\caption{Dynamical observables for $L = 6$.
(a): Spin auto-correlation function. (b): Typical relaxation time of spin variables.
(c): Orbital auto-correlation function. (d): Typical relaxation time of orbital variables.
The black solid lines in the figure (b) and (d) denote the fits by $\tau = A(T-T_{\rm c})^{-z\nu}$ with $(A, T_{\rm c}, z\nu)_s = (3.99(89) , 0.0695(19), 2.79(12))$ and $(A, T_{\rm c}, z\nu)_\sigma= (4.27(31), 0.0686(5), 3.55(4))$ by using the least squares method.
Insets: Scaling plots of auto-correlation functions using their relaxation times. 
The solid lines represent the exponential fits.
}
\label{Fig2}
\end{figure}

To explore spin and orbital freezing we examined the auto-correlation functions (ACFs) for both degrees of freedom
$C_{s}(t) = N^{-1}\sum_{i = 1}^N \bm {S}_i(t)\cdot\bm{S}_i(0)$ and $C_{\sigma}(t) = N^{-1}\sum_{i = 1}^N \bm{\sigma}_i(t) \cdot \bm{\sigma}_i(0)$
measured in equilibrium.
The behavior of the spin and orbital ACFs are shown in Figs. \ref{Fig2}(a) and \ref{Fig2}(c).
Both ACFs 
indicate the slowing down of the dynamics. 
As a quantitative measure of dynamics, 
we define the relaxation times $\tau_s(T)$ and $\tau_\sigma(T)$ for each degree of freedom 
as the time-scale such that the ACFs decrease down to 0.5, 
which are evaluated as shown in Figs. 2(b) and (d).
Both relaxation times diverge at the same temperature $T_c \approx 0.07$ with power laws 
although their exponents differ. 
This suggests that the spin and orbital are frozen simultaneously. 
We checked that there is no finite size effect within the time scale which we analyzed. \blue{We also analyzed data at $\epsilon=0.65$ and obtained
consistent results. (See SI).}
In the insets of Figs. \ref{Fig2}(b) and \ref{Fig2}(d) 
we show the scaling plots of ACFs assuming a scaling form, $C(t,T) = \tilde{C}(t/\tau(T))$, which is clearly satisfied.
The scaling functions turn out to be well fitted by simple exponentials.
We note that the relaxation function in the Edwards-Anderson models in the paramagnetic phase 
is more complicated  \cite{ogielski1985dynamics,yoshino1993monte}
presumably due to Griffith singularity  \cite{bray1988dynamics}.

\begin{figure}[t]
\includegraphics[clip,width=75mm]{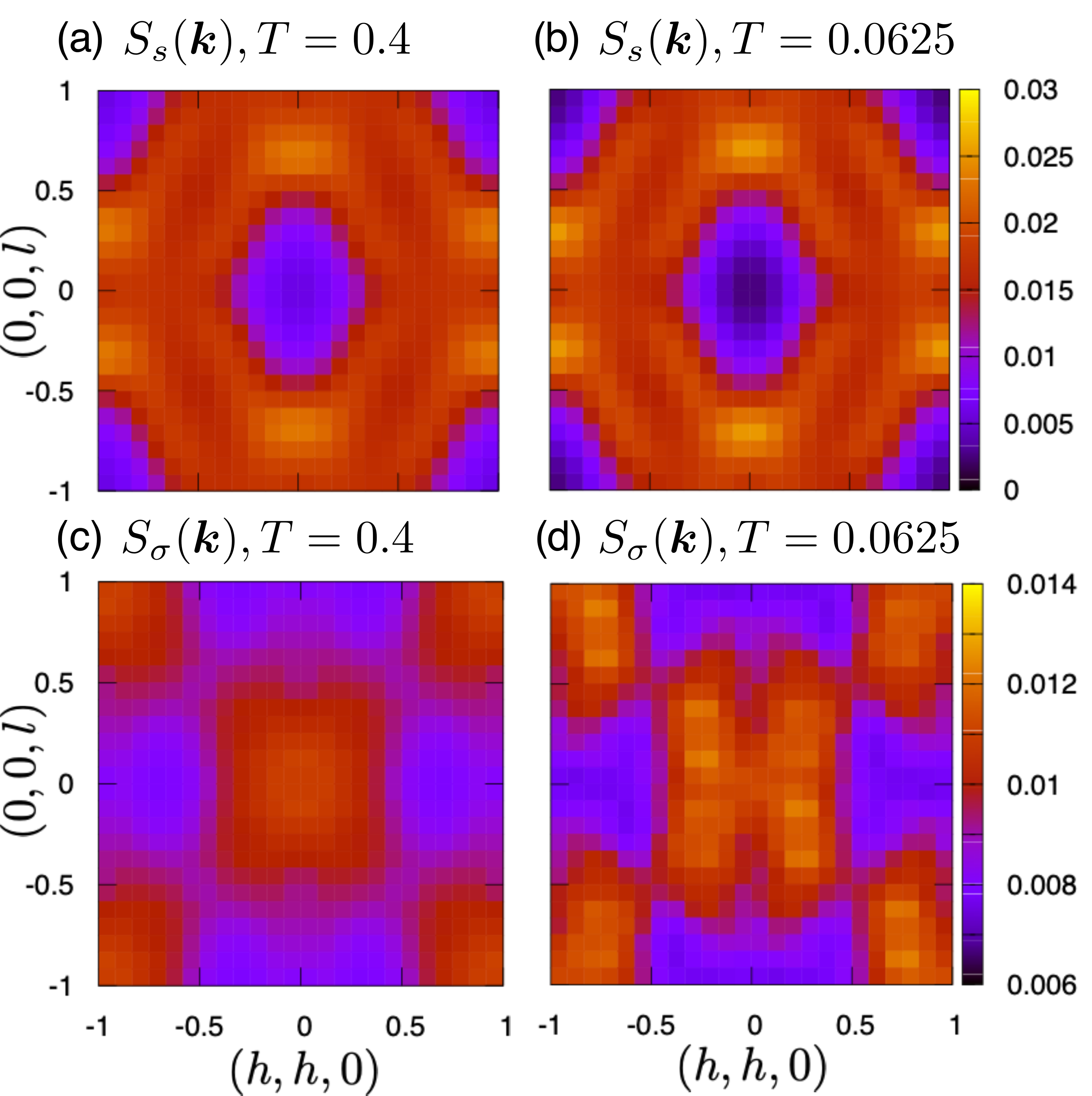}
\caption{ \red{Structure factors
of the spins (a),(b) and orbitals (lattice distortions) (c),(d)
on the $(h,h,l)/\pi$ plane for $L=6$. Panels (a) and (c) are for $T = 0.4$
while (b) and (d) are for $T = 0.0625$.
  Pinch points can be seen, for example at $(1/2,1/2,1/2)$, at low temperatures.
  }
}
\label{Fig3}
\end{figure}

To get further insights we analyzed
the static-structure factors of spins and lattice distortions defined 
respectively as 
$S_{s}(\bm{k}) = N^{-1} | \sum_{i = 1}^N e^{-i\bm{k}\cdot\bm{r}_i} \bm{S}_1\cdot\bm{S}_i|$ and
$S_{\sigma}(\bm{k}) = (4N)^{-1} | \sum_{i = 1}^4\sum_{j = 1}^N e^{-i\bm{k}\cdot(\bm{r}_j-\bm{r}_i)}\bm{\sigma}_i\cdot\bm{\sigma}_j |$, 
where $\bm{k}$ 
is a wave vector. 
In figures \ref{Fig3}(a) and \ref{Fig3}(b) we show $S_{s}(h,h,l)$ measured
above and below $T_c$, which show  no notable differences.
No Bragg peaks are observed bellow $T_c$, indicating that there is no magnetic long range order.
In Figs. \ref{Fig3}(c) and \ref{Fig3}(d), 
we also find no sign of long range order in the lattice distortions.
\red{The pinch-points similar to those observed in spin-ice systems
 \cite{bramwell2001spin}\footnote{
  Note however that here we don't multiply any scattering factor $|f(\bm{k})|$
  to the structure factor.
} can be seen.
The pinch-points reflect the fraction of tetrahedra that satisfy the ice-rule  which increases smoothly with decreasing temperature without any anomaly at
the transition temperature $T_{\rm c}$. (see SI).}

The present SG transition can be detected by the two thermodynamic quantities. 
One is the nonlinear magnetic susceptibility, 
\begin{equation}
\chi_3 = \eval{ \frac{1}{3}\sum_\mu \frac{\partial^3 \langle m_\mu \rangle_{\rm eq}}{\partial h_\mu^3}}_{h_\mu \to 0},
\end{equation} 
where $m_\mu = N^{-1}\sum_{i = 1}^N S_{i\mu}$ and $h_\mu$ are the magnetization and the magnetic field, respectively, 
in the $\mu$-direction ($\mu = x, y, z$). 
The other is the nonlinear dielectric susceptibility, 
\begin{equation}
\chi^\sigma_3 = \eval{ \frac{1}{4}\sum_{\nu} \frac{\partial^3 \langle p_\nu \rangle_{\rm eq}}{\partial E_\nu^3}}_{E_\nu \to 0},
\end{equation}
where $p_\nu = N^{-1}\sum_{i = 1}^N \bm{\sigma}_i \cdot \hat{\bm{e}}_\nu$ 
is the dielectric polarization in the $\hat{\bm{e}}_\nu$-direction and $E_\nu$ is the electric field. 
Fortunately, the since Mo$^{4+}$ ion has an electric charge, the fluctuation of the orbital glass ordering can be detected electrically. 
Thus, the two quantities are measurable in laboratories and allow a direct comparison between theories and experiments. 
In conventional theories \cite{mezard1987spin,fisher1988equilibrium},
one often analyzes the SG susceptibility, $\chi_{\rm SG}$, which is proportional to $\chi_3$. 
Although $\chi_{\rm SG}$ is much easier to calculate with better accuracy in the presence of quenched disorder 
by using the overlap between two independent replicas, 
it is not convenient for our model, since the Hamiltonian is translationally invariant
so that the replicas usually do not overlap. 

\begin{figure}[t]
\includegraphics[clip,width=85mm]{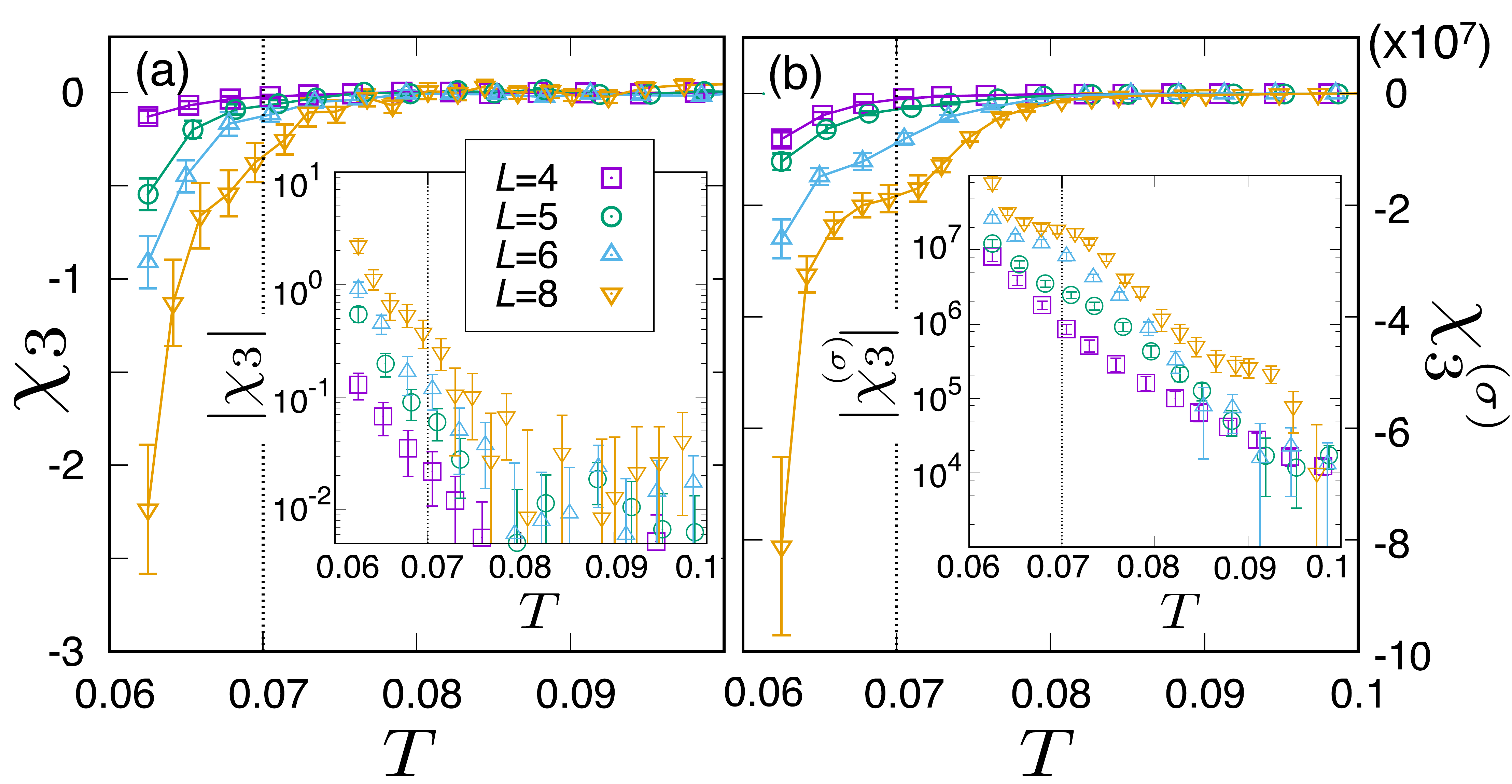}
\caption{
(a), (b): Size dependence of nonlinear magnetic and dielectric susceptibilities (Insets: logarithmic scale of absolute values).
The dashed line represents the freezing temperature $T_{\rm c} \approx 0.07$ determined by the auto-correlation function. 
}
\label{Fig4}
\end{figure}

Figure \ref{Fig4}(a) shows $\chi_3$ as a function of temperature for several system size $L$. 
In overall, it takes the negative value, and from slightly above $T_c$, 
its amplitude starts to grow rapidly in lowering the temperature. 
This growth is enhanced for larger $L$, strongly suggesting that the divergence of $\chi_3$ remains in the thermodynamic limit, 
which is consistent with the experimental observation in Y$_2$Mo$_2$O$_7$. 
Similar behavior is observed for $\chi_3^{(\sigma)}$, indicating the freezing of the orbital 
degrees of freedom into a disordered state. 

Let us discuss the relevance of our results with experiments. 
The existence of the SG transition itself is already established in Y$_2$Mo$_2$O$_7$ and its families, 
and although its origin has not been understood, a recent analysis showed that the lattice distortions possibly plays 
a certain role in the glass transition. 
Our model is constructed based on this finding by including the direct coupling between 
the lattice distortion and the spin. 
The two degrees of freedom simultaneously undergo a glass transition, indicating that 
they generate an effect of dynamical randomness to each other. 
Experiments can test our scenario by examining 
whether the nonlinear dielectric susceptibility diverges in the vicinity of $T_{\rm c}$, 
where already the nonlinear magnetic susceptibility shows the divergence. 
Previously, the orbital-glass transition was observed in the FeCr$_2$S$_4$, 
where the dielectric spectroscopy measurement played a major role, detecting the slowing down of orbital dynamics
 \cite{fichtl2005orbital}. Similar techniques shall be applied to the pyrochlore materials
\red{and in addition to that the ZFC/FC anomaly \cite{gingras1997static}
 shall be tested in linear dielectric response of the pyrhochlore materials.}

Let us finally comment on the scenario often posed experimentally, which expects
the freezing of the orbital degrees of freedom at a higher temperature than the SG transition   \cite{booth2000local,thygesen2017orbital}. 
Once the orbital configuration is frozen, 
our model is reduced to the standard EA model of Heisenberg spins 
with quenched randomness on a pyrochlore lattice, which is known to exhibit a SG transition \cite{saunders2007spin,shinaoka2011spin}.  
Actually, if we increase $\epsilon$ in Eq.(\ref{hamiltonian}), 
the ice-rule configuration of orbitals is selected at the higher temperature, 
which is detected by the broad peak in the heat capacity(see SI). 
However, this peak does not mean the transition but a crossover;
Below that temperature, the orbitals are still dynamically fluctuating among huge numbers of 
quasi-degenerate ice-rule configurations. 
In lowering the temperature, the orbital dynamics naturally slows down but without any thermodynamic anomaly
much as usual spin ices  \cite{jaubert2009signature}.
Interestingly, at a lower temperature the heat capacity shows a second peak (see SI)
which should be attributed to coupling between the spin and lattice distortions. 
In our scenario, we anticipate the thermodynamic spin-orbital glass transition at around that temperature. 

Some other effect such as the spin-orbit interaction
 \cite{shinaoka2013spin}, the spin-phonon coupling and 
longer range interactions, which are not included in our model,
may be needed
to reproduce more precisely the experimental measurements, 
whereas our findings proved that the interplay of nearest-neighbor interactions and the dynamical JT distortions 
can solely drive the system to the glass transition. 
The present model can be further applied to other SG frustrated magnets such as Tb$_2$Mo$_2$O$_7$ whose magnetic ion is JT active.

To summarize, we introduced the realistic model without quenched randomness 
consisting of spin and orbital degrees of freedom, that shows the 
thermodynamic glass transition for the first time in the finite dimensional periodic lattice. 
We performed the dynamical simulations in the equilibrium and found that 
the relaxation times of both spins and orbitals show a power-law divergence 
at the same temperature, $T_{\rm c}$.
The nonlinear magnetic susceptibility and the nonlinear dielectric susceptibility together show 
negative divergence at around $T_{\rm c}$, signaling the simultaneous glass transition of these two degrees of freedom. 
The former is consistent with the experiments already performed in the pyrochlore oxide, Y$_2$Mo$_2$O$_7$, 
whereas the latter can be tested in the near future to fix the long standing problem on the origin of the SG transition 
in this disorder free crystalline solid.

\begin{acknowledgments}
This work was supported by KAKENHI (No. 19H01812, 17K05533, 17K05497,18H01173,17H02916) from MEXT, Japan.
The computation in this work has been done using the facilities of the Supercomputer Center, the Institute for Solid State Physics, the University of Tokyo, OCTOPUS and supercomputer system SX-ACE at the Cybermedia Center, Osaka University.
\end{acknowledgments}

\bibliographystyle{apsrev4-1}
\include{y2mo2o7_prl_v9.2.bbl}

\onecolumngrid
\appendix
\section{ Microscopic justification of the effective model}

\begin{figure*}[b]
\begin{tabular}{c}
\begin{minipage}{0.4\textwidth}
\centering
  \begin{tabular}{|l|c|c|c|}
  \hline
     & $\alpha = 116^\circ$ & $\alpha = 127^\circ$ & $\alpha = 139^\circ$ \\ \hline \hline
    $|t_{p(X),a_{1g}}|^2$ & 0.951913 & 0.561214 & 0.151767 \\ \hline
    $|t_{p(X),e_{g\pm}'}|^2$ & 1.97879 & 2.42526 & 2.69398 \\ \hline
    $|t_{p(Z),a_{1g}}|^2$ & 0.0580811 & 0.323847 & 0.738423 \\ \hline
    $|t_{p(Z),e_{g\pm}'}|^2$ & 0.97992 & 0.572566 & 0.216754 \\ \hline
  \end{tabular}
\tblcaption{The square of the transfer integral $|t|^2$ (eV$^2$) for three angles $\alpha = 116^\circ, 127^\circ, 139^\circ$ evaluated by the Slater-Koster parameter.}
\label{transfer_a1g_eg}
\end{minipage}
\hspace{1cm}
\begin{minipage}{0.45\textwidth}
\centering
\includegraphics[clip,width=60mm]{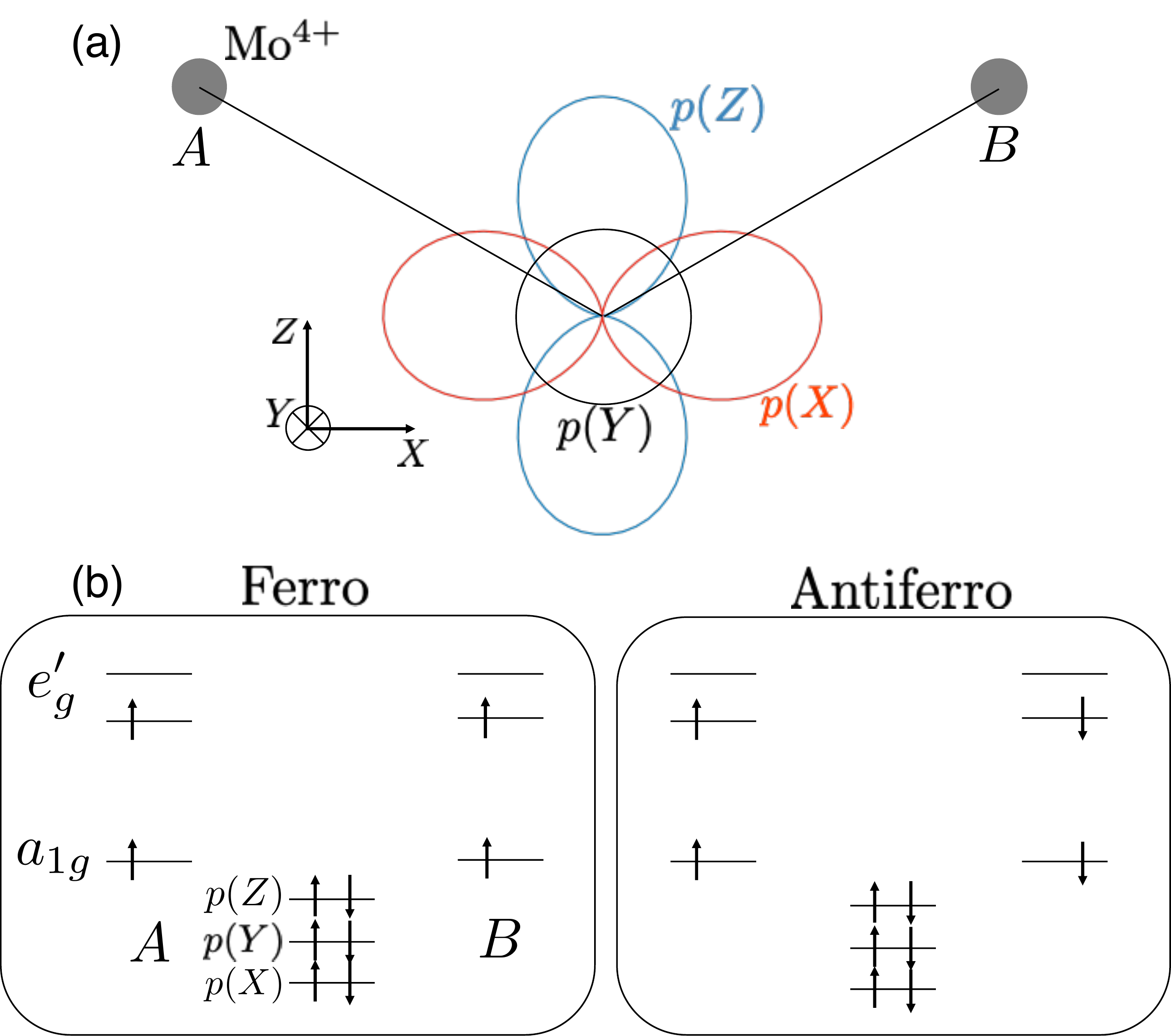}
\figcaption{
(a) Relative positions between two Mo$^{+4}$ ions (A,B) and O(2p) orbitals. 
Since one can construct three degenerate $p$-orbitals in an arbitrary combination, 
we define the new orbitals $\ket{p(X)}$, $\ket{p(Y)}$, $\ket{p(Z)}$ in a symmetric manner in terms of A and B. 
(b): Electronic configurations of ferromagnetic and antiferromagnetic low energy Mott insulating states 
in the strong coupling limit of Eq.(S1). 
}
\label{porbital}
\end{minipage}
\end{tabular}
\end{figure*}

\red{
In a pyrochlore oxide Y$_2$Mo$_2$O$_7$, magnetic ion Mo$^{4+}(4d^2)$ is affected by trigonal crystal field, 
and its triply $t_{2g}$ orbitals split into a single $a_{1g}$ orbital and doubly $e_g'$ orbitals, 
where $a_{1g}$ has the lower energy than $e_g'$. 
The spins of two electrons on Mo$^{4+}$ form an effective $S=1$ following Hund's rule.
This is because the magnitude of energy splitting between $a_{1g}$ and $e_g'$ is smaller than the strength of Hund's coupling. \cite{solovyev2003effects}
We further take account of the Jahn-Teller lattice distortion, which is observed experimentally \cite{booth2000local,thygesen2017orbital}.
Ref.\cite{thygesen2017orbital} shows that each Mo$^{4+}$ moves away from the center of the tetrahedron, 
and accordingly, the Mo-O-Mo angle varies from its equilibrium position $\alpha=$127$^\circ$ to 116$^\circ$ for the in-in bond 
and 139$^\circ$ for the out-out bond. 
We checked that the change of the angle is large enough to give the different sign of the exchange interaction as in the following.
}

\red{
  We consider an extended Kanamori Hamiltonian consisting of $d$-orbitals $a = \ket{a_{1g}}, \ket{e_{g+}'}, \ket{e_{g-}'}$ 
of Mo$^{4+}$ ions and $p$-orbitals $b = \ket{p(X)},\ket{p(Y)},\ket{p(Z)}$ on the O$^{2-}$ ion. 
As shown in Fig. S\ref{porbital}, we consider two adjacent Mo$^{4+}$ ions, A and B, and the relative position 
of the degenerate three $p$-orbitals whose quantization axis is defined as such that it is line symmetric with respect to  A and B. 
The Hamiltonian is given as 
\begin{align}
\nonumber
{\mathcal H} = &-\sum_{i = {\rm A},{\rm B}}\sum_{a,b}\sum_{\sigma} t_{a,b} \qty(c_{i a \sigma}^\dagger c_{o b \sigma} + c_{o b \sigma}^\dagger c_{i a \sigma}) + \sum_{i = {\rm A},{\rm B}} \sum_{a}\Delta e_a \sum_{\sigma} c_{i a \sigma}^\dagger c_{i a \sigma} \\
&+ \sum_{i = {\rm A},{\rm B}} \sum_{a,a'} \sum_{\sigma,\sigma'} \frac{K(a,a') - \delta_{\sigma,\sigma'}J(a,a')}{2} c_{i a \sigma}^\dagger c_{i a \sigma}c_{i a' \sigma'}^\dagger c_{i a' \sigma'} + \frac{U_o}{2}\sum_{b} c_{o b \sigma}^\dagger c_{o b \sigma}c_{o b' \sigma'}^\dagger c_{o b' \sigma'},
\end{align}
where $c_{la\sigma}/c_{la\sigma}^\dagger$ are the annihilation/creation operators 
of the $a$-orbital of either the Mo-site ($l={\rm A},{\rm B}$) or O-site ($l=o$) with spin $\sigma$. 
The transfer integral $t_{a,b}$ in the first term are those between $a$-orbital of Mo$(4d)$ and $b$-orbital of O$(2p)$, 
which depends on both the distances between ions and the Mo-O-Mo angle $\alpha$ \cite{harrison2012electronic}. 
We calculated $t_{a,b}$ based on the Slater-Koster parameter, as presented in Table. \ref{transfer_a1g_eg}.
The hopping between $p(Y)$ and any Mo$(4d)$ orbitals are relatively small and does not depend much on $\alpha$, thus are neglected.
}

\red{
The second term represents the excitation energy from O$(2p)$ to Mo$(4d)$ and we adopt 
$\Delta e_{a_{1g}} = 2.0$ and $\Delta e_{e_{g\pm}'} = 2.5$ taken from the first principles band calculation in Ref.\cite{solovyev2003effects}.
The third and fourth terms are the on-site Coulomb interactions on Mo$^{4+}$ and O$^{2-}$, respectively.
$K(\alpha,\alpha')$ and $J(\alpha,\alpha')$ are the Coulomb and exchange integrations 
which are described by the Slater-Condon parameters $F^0,F^2$ and $F^4$. 
We adopt $F^2 = 4.52$ and $F^4/F^2 = 0.63$ \cite{solovyev2003effects}.
We leave $F^0=U_{\rm Mo}$ and $U_{o}$, which represent the strength of
the on-site Coulomb interactions on the Mo and O ions respectively,
as parameters.
}

\red{
We start from the strong coupling limit, $t_{a,b}\ll U_{\rm Mo}, U_{o}, \Delta e_{a}$, 
where the Mott insulating ground state consists of fully occupied O$(2p)$ orbitals and the 
${\rm A/B}$ ions each having two electrons that occupy $a_{1g}$ and either of $e_{g+}'$/$e_{g-}'$ 
(See Fig. S1 (a)). 
Since these two electrons on the same Mo$(4d)$ ions form the $S=1$ triplet due to Hund's coupling, 
there are four different spin configurations $(S^z_{\rm A},S^z_{\rm B})=(\uparrow,\uparrow), (\uparrow,\downarrow), (\downarrow,\uparrow), (\downarrow,\downarrow)$ which we take as a low energy basis. 
The leading term which gives the effective exchange interaction $J_{\rm eff}\bm{S}_{\rm A}\cdot\bm{S}_{\rm B}$ is obtained at the fourth of perturbation in terms of $t_{a,b}$. 
By considering all the perturbation processes of either of the two electrons on the $p$-oribals each to hop to 
the empty orbitals on A and B ions and coming back, one can evaluate the energy gain $E_{\rm F/AF}^{(4)}$ for 
the parallel and anti-parallel spin orientations of A and B ions. 
The effective exchange interaction is given by $J_{\rm eff} = (E^{(4)}_{\rm F} - E^{(4)}_{\rm AF})/3$, where denominator 3 is from the effective spin $S = 1$. 
The fact that both $|t_{p(X),a_{1g}}|^2$ and $|t_{p(Z),e_{g\pm}'}|^2$ vary significantly with $\alpha$ 
affects the relative energy gain of $E^{(4)}_{\rm F}$ and $E^{(4)}_{\rm AF}$, the sign of $J_{\rm eff}$ also changes 
its sign at $\alpha \sim 127^\circ$, as we present in Fig. 1(d) in the main text.
}

\vspace*{1cm}
\section{Details of the Monte Carlo simulation}

Here we explain the details of our Monte Carlo simulation.
For updating lattice displacements $\sigma_i$ $(i = 1, 2, 3, ..., N )$ we used the conventional single-spin-flip Metropolis method.
However, it is hard to investigate the thermodynamic properties of the ice-type system at low temperature due to multiple energy minima with high energy barrier.
Therefore, we also adopted a nonlocal update method called the loop algorithm. \cite{melko2001long}
The nonlocal update consists of two steps:
(1) We look for the loop consisting of only \textit{in}-\textit{out} bonds.
(2) We flip all the lattice displacements on the loop simultaneously with the acceptance rate determined by the Metropolis rule.
Note that the
energy contribution  from the second term of the Hamiltonian remains unchanged in this update.
For updating Heisenberg spins we used the Metropolis Reflection method \cite{pawig1998monte} and the over-relaxation method. \cite{PhysRevB.53.2537}
In the Metropolis Reflection method, we prepare a plane dividing spin space into two equal parts randomly.
To update a spin $\bm{S}_i$, a new candidate spin configuration $\bm{S}_i^\prime$ is created by reflecting $\bm{S}_i$ by the plane. 
More precisely the sign of the perpendicular component to plane is changed. 
The reflection is performed with probability determined by Metropolis rule. 
In a unit Monte Carlo step (MCS), we try the update for all spins  $\bm{S}_i$ $(i = 1, 2, 3, ..., N )$ and apply the same plane to reflect all spins. 
In the over-relaxation method, we first calculate the effective magnetic field $\bm{h}_{\rm eff}$ around a spin $\bm{S}_i$, and then rotate the spin around $\bm{h}_{\rm eff}$ by an angle of $\pi$.
A unit Monte Carlo step is consisted of the following steps.
\begin{enumerate}
\item Sequential single-spin flip for all lattice displacements with the conventional Metropolis method
\item $L$ times loop update for lattice displacements
\item Calculating all the coupling constants $J_{\bm{\sigma}_i,\bm{\sigma}_j}$
\item Sequential single-spin flip for all Heisenberg spins with the Metropolis Reflection method
\item $L$ times sequential single-spin flip for all Heisenberg spins with the over-relaxation method
\end{enumerate}
Here $L$ is the system size defined in the main text.

The equilibrium states were realized by the replica exchange method. \cite{hukushima1996exchange, hukushima1999domain}
We prepared 48 replicas for $L = 4,5,6$ and 72 replicas for $L = 8$. 
We determined the maximum temperature and the minimum temperature as $T_{\rm max} = 0.400$ and $T_{\rm min} = 0.0625$ respectively.
We optimized the temperature set so that the exchange rate is independent of $T$. \cite{hukushima1999domain}
We performed the trial to exchange the replicas with the interval 15 (MCS).
Initial spin and lattice configuration for each replica were prepared a fully random configuration.
MCSs for thermalization was determined as $3.0 \times 10^{7}$ (MCS) for all system sizes.
Observations of physical quantities are made during additional $3.0 \times 10^7$ (MCS) to evaluate their thermal averages $\langle...\rangle_{\rm eq}$ by their MCS averages.

In dynamical simulations in equilibrium, the initial configurations ($t = 0$ (MCS)) are created by above equilibration.
In order to observe the natural time evolution we apply only the sequential single-spin flip for both degrees of freedom.
A unit Monte Carlo step is consisted of the following steps.
\begin{enumerate}
\item Sequential single-spin flip for all lattice displacements with the conventional Metropolis method
\item Calculating all the coupling constants $J_{\bm{\sigma}_i,\bm{\sigma}_j}$
\item Sequential single-spin flip for all Heisenberg spins with the Metropolis Reflection method
\end{enumerate}

\section{Definitions of the non-linear magnetic and dielectric
  susceptibilities}

Here we present the detailed definition of the non-linear susceptibilities
we measured in our simulations.
From fluctuation formulae we obtain the non-linear magnetic susceptibility as
\begin{align}
\nonumber
\chi_3 &= \frac{1}{3}\sum_{\mu} \frac{\partial^3 \expval{ m_\mu }_{\rm eq}}{\partial h_\mu^3} \\
 &= \frac{\beta^3 N^3}{3}\sum_{\mu} \Bigl(\expval{m_\mu^4}_{\rm eq} - 4\expval{m_\mu}_{\rm eq}\expval{m_\mu^3}_{\rm eq}-3\expval{m_\mu^2}_{\rm eq}^2 + 12\expval{m_\mu^2}_{\rm eq}\expval{m_\mu}_{\rm eq}^2 - 6\expval{m_\mu}_{\rm eq}^4 \Bigr).
\end{align}
Within our Monte Carlo simulations of finite sized systems,
odd order moments of the magnetization vanish
due to the rotational symmetry, e.g. $\expval{m_\mu}_{\rm eq}=\expval{m_\mu^3}_{\rm eq}=0$.
Hence we can evaluate it simply as,
\begin{align}
\chi_3 = \frac{\beta^3 N^3}{3}\sum_{\mu} \Bigl(\expval{m_\mu^4}_{\rm eq} - 3\expval{m_\mu^2}_{\rm eq}^2 \Bigr).
\end{align}

Similarly we evaluate the the non-linear dielectric susceptibility in the same manner as,
\begin{align}
\chi^{(\sigma)}_3 = \frac{\beta^3 N^3}{4}\sum_{\nu} \Bigl(\expval{p_\nu^4}_{\rm eq} - 3\expval{p_\nu^2}_{\rm eq}^2 \Bigr).
\end{align}

\afterpage{\clearpage}
\newpage

\section{Spin and orbital time auto-correlation functions}

In main text, we present the results of the time auto-correlation functions
of the spin and orbital
at $\epsilon = 0.60$. The data suggest that spin and orbital degrees of freedom simultaneously freeze at $T_{\rm c} \approx 0.07$.
Here we show an additional data set obtained at $\epsilon = 0.65$.
The scaling analysis of the data suggests
a simultaneous glass transitions at $T_{\rm c} \approx 0.086$.
Notably, the exponents $z\nu$ agree with those obtained
at $\epsilon = 0.6$ within the numerical accuracy, suggesting a universality.

\begin{figure*}[h]
\includegraphics[clip,width=130mm]{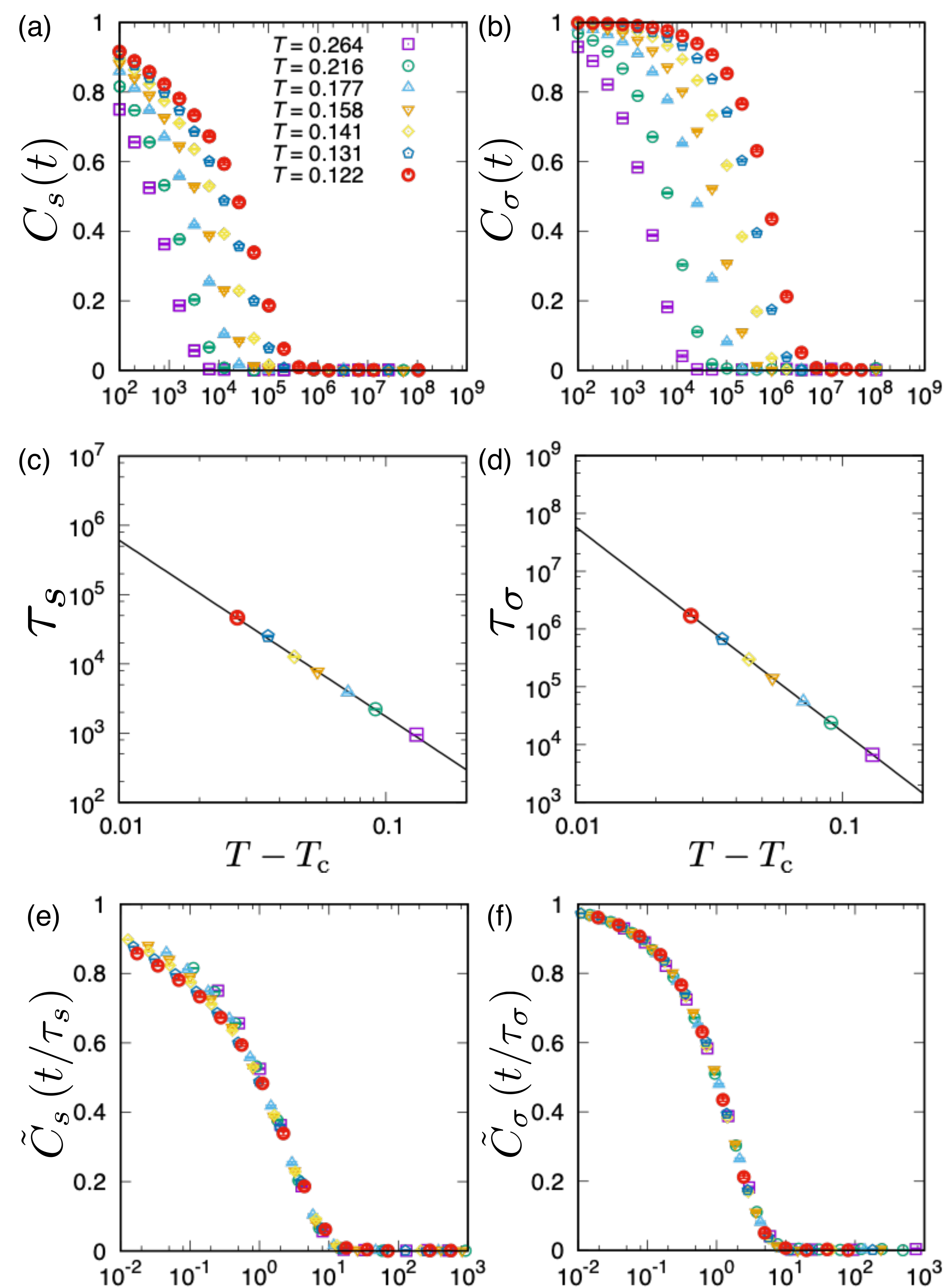}
\caption{
Dynamical observables for $L = 5,~\epsilon = 0.65$.
(a): Spin auto-correlation function.
(b): Orbital auto-correlation function.
(c): Typical relaxation time of spin variables.
(d): Typical relaxation time of orbital variables.
The black lines in the figure (a) and (b) denote the fits by $\tau = A(T-T_{\rm c})^{-z\nu}$ with $(A,T_{\rm c},z\nu)_s = (4.91(154), 0.0855(26), 2.55(15))$ and $(A,T_{\rm c},z\nu)_\sigma = (4.78(21), 0.0862(5), 3.54(3))$ by using the least squares method.
(e), (f): Scaling plots of spin and orbital auto-correlation function, respectively.
}
\label{relaxation}
\end{figure*}

\afterpage{\clearpage}
\newpage

\section{Heat capacity}

We show below
the temperature dependence of heat capacity for various $\epsilon$.
We can see a broad peak in the vicinity of spin glass transition temperature $T_{\rm c}$, similarly to the observations in usual spinglass systems
(for example see Chap 3.1.2 of \cite{mydosh2014spin})
and the pyrochlore SG \cite{raju1992magnetic}.
One see that 
the dependence on the system size is very weak (see Fig. S3 (a)).

For larger $\epsilon$, another broad peak emerges at higher temperatures
(see Fig. S3 (c)), which
can be interpreted as a reflection of the crossover
from purely random phase to ice like phase.
(see Fig. S4.)

\begin{figure*}[h]
\includegraphics[clip,width=130mm]{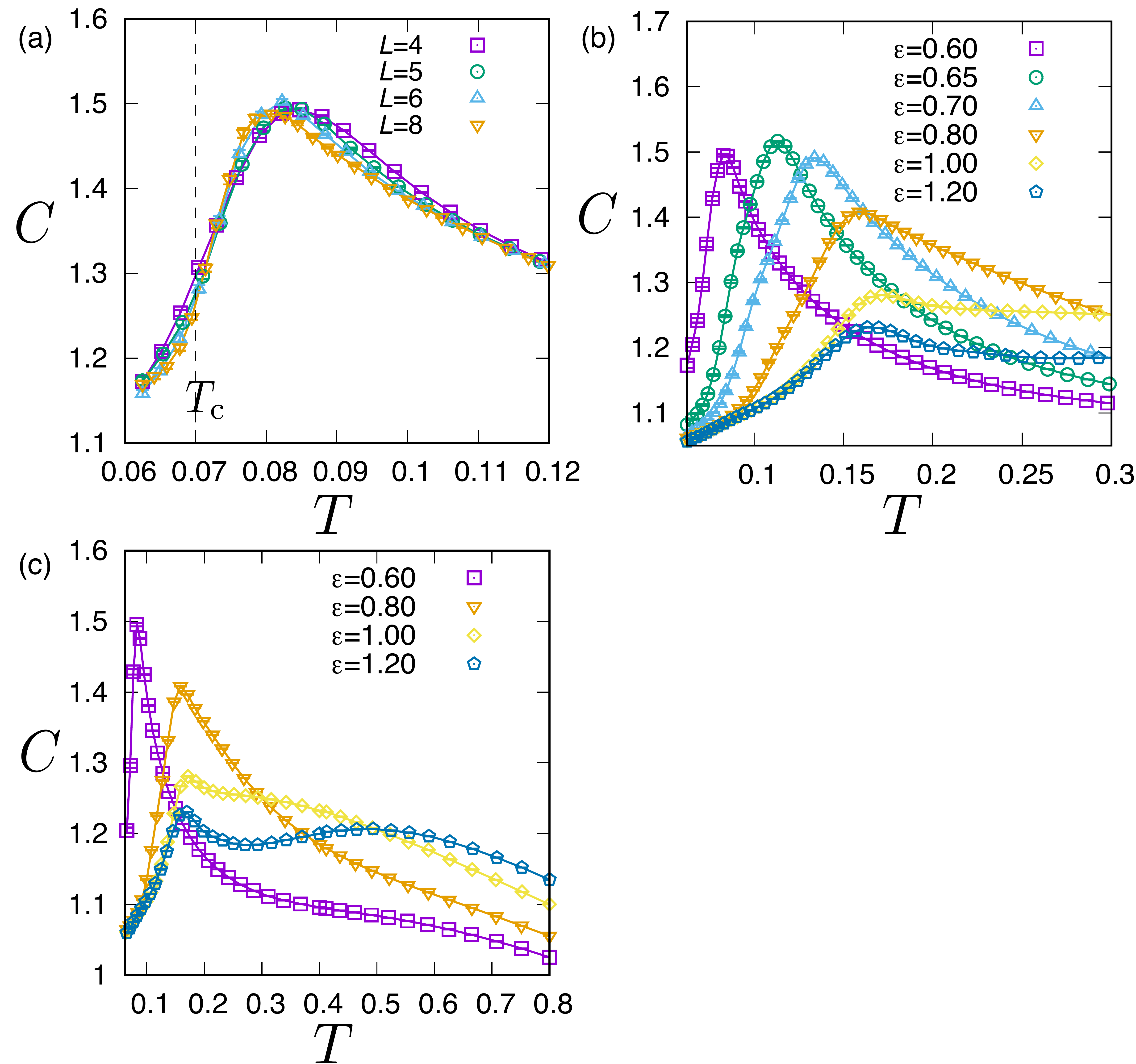}
\caption{
Heat capacity calculated as $C = N^{-1}\beta^2 \qty(\expval{E^2}_{\rm eq} - \expval{E}_{\rm eq}^2)$.
(a): Size dependence at $\epsilon = 0.6$.
(b): $\epsilon$ dependence for $L=5$ in the narrow range of $T=0.0625$ to $T = 0.3$.
(c): $\epsilon$ dependence for $L=5$ in the wide range of $T=0.0625$ to $T = 0.8$.
}
\label{heatcap}
\end{figure*}

\afterpage{\clearpage}
\newpage

\section{Fraction of ice like lattice distortions}

Fig. S4 shows the temperature-dependence of the fraction of ice like
2-in 2-out structure of the lattice distortions for various $\epsilon$.
 It can be seen that the onset of ice like structure 
 is gradual. The 2ndary peak of the heat-capacity observed
 at large $\epsilon$ (See Fig. S3(c)) can be interpreted as
due to the onset of the ice like structures.

\begin{figure*}[h]
\includegraphics[clip,width=130mm]{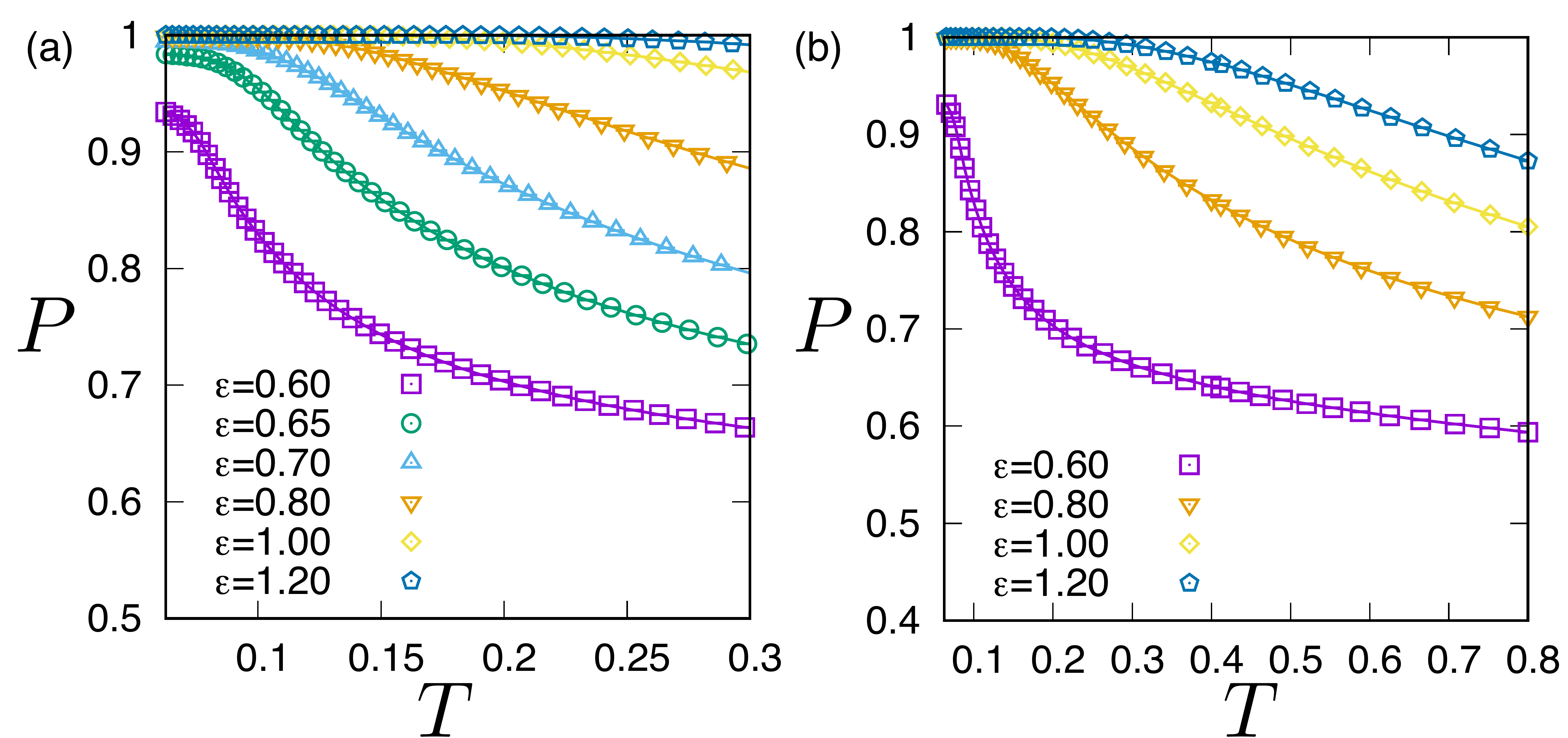}
\caption{
Fraction of the ice structures given by $N_\text{2-\textit{in}-2-\textit{out}}/N_{\triangle}$.
Here $N_{\triangle} = N_{\rm spin}/2$ represents the number of all tetrahedra and $N_\text{2-\textit{in}-2-out}$ represent the number of tetrahedra which are distorted into 2-\textit{in}-2-\textit{out} structures.
(a): $\epsilon$ dependence for $L=5$ in the narrow range of $T=0.0625$ to $T = 0.3$.
(b): $\epsilon$ dependence for $L=5$ in the wide range of $T=0.0625$ to $T = 0.8$.
}
\label{fraction}
\end{figure*}

\section{Non-linear magnetic and dielectric susceptibilities}

In the main text, we present the temperature dependence of magnetic and dielectric non-linear susceptibilities for $\epsilon = 0.60$.
Here, we show them for various $\epsilon$.

\begin{figure*}[h]
\includegraphics[clip,width=130mm]{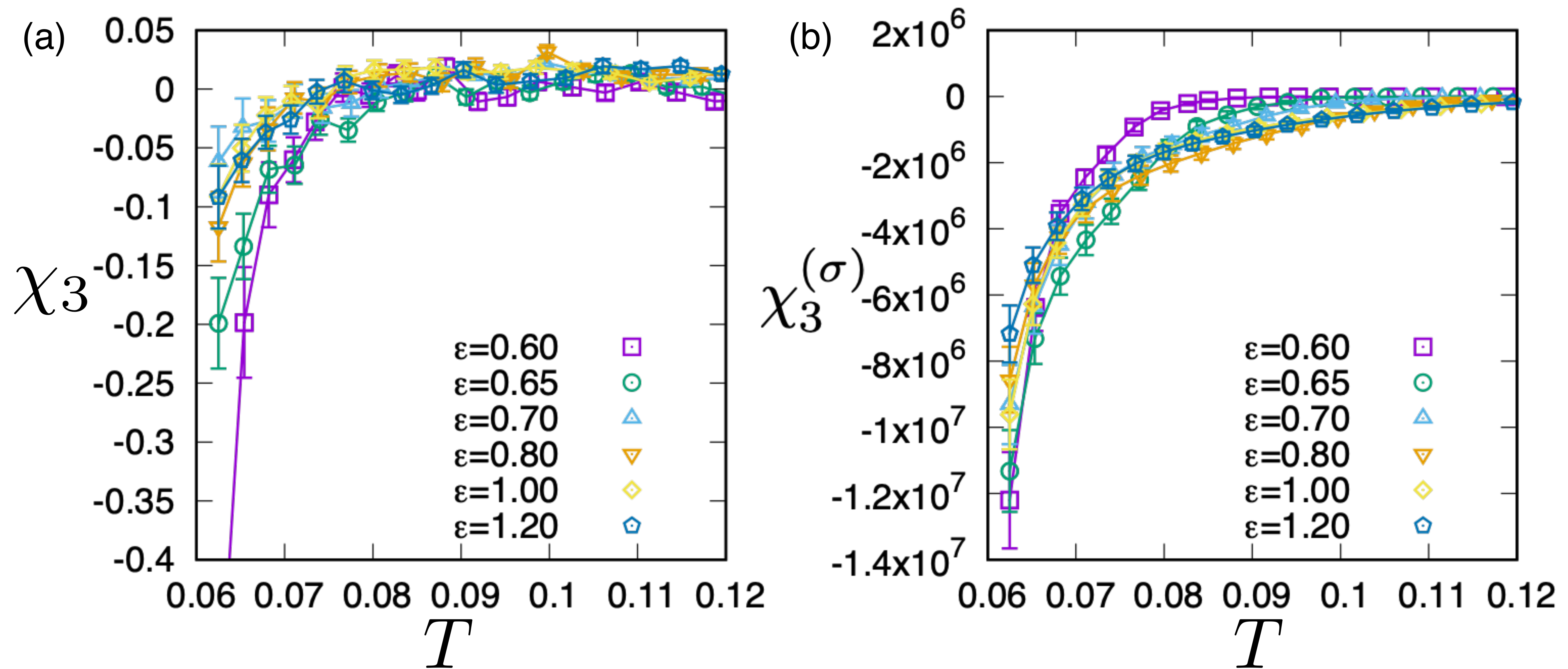}
\caption{
(a): $\epsilon$ dependence of the non-linear magnetic susceptibility for $L=5$.
(b): $\epsilon$ dependence of the non-linear dielectric susceptibility for $L=5$.
}
\label{suscep}
\end{figure*}

\end{document}